# LINEAR LATTICE AND TRAJECTORY RECONSTRUCTION AND CORRECTION AT FAST LINEAR ACCELERATOR

A. Romanov[†], D. Edstrom, Fermilab, Batavia, IL
A. Halavanau, Northern Illinois University, DeKalb, IL

*Abstract*

The low energy part of the FAST linear accelerator based on 1.3 GHz superconducting RF cavities was successfully commissioned [1]. During commissioning, beam based model dependent methods were used to correct linear lattice and trajectory. Lattice correction algorithm is based on analysis of beam shape from profile monitors and trajectory responses to dipole correctors. Trajectory responses to field gradient variations in quadrupoles and phase variations in superconducting RF cavities were used to correct bunch offsets in quadrupoles and accelerating cavities relative to their magnetic axes. Details of used methods and experimental results are presented.

## INTRODUCTION

Many of the experiments planned at the FAST linear accelerator require a beam with low emittances and well-known beam parameters along the line. Proper element positioning and calibration ensures good experimental lattice agreement with the model, but in order to finely tune the lattice parameters, beam-based methods are necessary. For example, steering the beam trajectory as close as possible to the magnetic axis of the elements not only gives better emittances, but also minimizes beam movements caused by changes in element parameters, coupled to their focusing strengths. After achieving the desired trajectory configuration, responses of the beam positions at BPMs to the corrector variations coupled with beam size analysis can give precisely tuned element parameters and determined initial conditions of the beam which allows to have relevant model of particles' distribution along the accelerator.

## METHOD

Trajectory and lattice corrections consist of two independent tasks. The first task is to find the difference between the real configuration and the model. The second is to find an optimal compensation schema. Both of these tasks can be formulated as inverse problems, where the goal is to find model parameters $p_i$ that minimize the difference between some experimental data $V_{exp}$ and the same values calculated from the model $V_{mod}=M(p_i)$.

To find trajectory position relative to the magnetic axes of the focusing elements, it is possible to use trajectory responses to variations in the focusing strength as $V_{exp}$ and parameters $p_i$ are relative coordinates of the trajectory. For short elements only two coordinates can be found. For long elements with significant variation of trajectory inside the element, such as strong solenoids or strings of RF cavities, it is possible to reconstruct both the position and the angle. At the second step, found trajectory distortions act as experimental data $V_{exp}$ and parameters $p_i$ are strengths of the correctors.

Lattice reconstruction is a more difficult task because of the larger amount of experimental data and number of lattice parameters involved. The following experimental data was used for the FAST lattice analysis:
- Trajectory responses to dipole correctors
- Beam second moments measured along the line

The following variable parameters of the FAST model were used:
- Gradients of magnetic fields in quadrupoles
- RF voltages in capture cavities
- RF phases in capture cavities
- Calibrations of correctors
- Calibrations of BPMs

Corrections of the lattice were straightforward, since parameters of quadrupoles and capture cavities can be adjusted individually.

## DETERMINING INITIAL CONDITIONS

The minimal normalized emittances of the beam are determined by the gun configuration and can't be decreased without damping, therefore it is important to have optimal gun parameters. In an uncoupled lattice, effective transverse emittances can be easily increased by a non-zero solenoidal field at the cathode [2].

Gun setup at the FAST linear accelerator, as well as at many others, has two solenoids: main solenoid for focusing and bucking solenoid for cancelling the solenoidal field at the cathode. Gun configuration at FAST has field values in iron yokes well below saturation level and by obtaining one setting with zero field, it is possible to derive favourable conditions for a wide range of settings by keeping the found proportion.

Consider a toy model where the focusing in the gun is determined by the main solenoid and the effective emittance is determined by the bucking solenoid. In this model, the smallest spot size at some screen separated from the gun only by the drift is achieved when the field at the cathode is equal to zero. FAST setup has X101 screen for the beam size measurements separated by about 90cm drift with the gun. Realistic ASTRA simulations of FAST for the beam size at the screen X101 during main and backing solenoids scan, shown on Figure 2, confirm toy model predictions. Figure 1 shows experimental dependence of the beam size on the currents in the main and the bucking solenoids and the point of minimal

† aromanov@fnal.gov

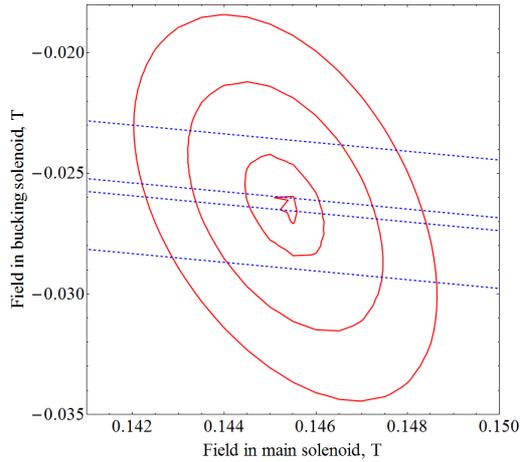

Figure 1: Experimental scan of size of the beam after the gun versus currents of main and bucking solenoids; red dot shows minimum of the smoothened surface

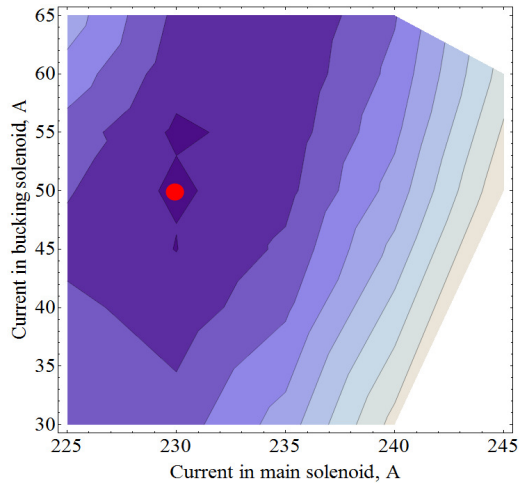

Figure 2: ASTRA simulations: red contours correspond to increase of the beam size at X101 by 1 %, 10 %, 50 % and 100 %; blue dashed lines indicate 1 % and 10 % of the residual field at cathode

size that corresponds to zero solenoidal field at the cathode.

In order to determine beam's Twiss parameters, it is often possible to use a simple geometrical method at some point along the lattice where two or more screens are available, separated only by drifts and with zero dispersion. The lattice should be adjusted so that the beam has minimum of the beta function at one screen, with corresponding beam size $\sigma_1$. Measurement of beam size at second screen $\sigma_2$ allows to calculate lattice parameters for a given configuration:

$$\varepsilon = \frac{\sigma_1 \sqrt{\sigma_2^2 - \sigma_1^2}}{L_{12}}; \quad \beta_1 = \frac{L_{12}}{\sqrt{(\sigma_2/\sigma_1)^2 - 1}}. \quad (1)$$

Here $\beta_1$ is beta-function at waist, $L_{12}$ is distance between screens. Method should be repeated for both planes.

With calibrated linear elements it is straightforward to back-trace obtained Twiss parameters to the beginning of the lattice for further lattice optimization.

## RESULTS

In order to determine emittances with geometrical method the focal point was made at the screen X120. Second sizes was measured at screens X109 located 652 cm upstream and X109 located 581 cm upstream. Table 1 contains resulted geometrical emitaces for a beam with momentum of about 40 MeV and negligible space charge effects. Statistical errors of size measurements are on the order of 1%, but systematical errors such as not perfect focusing at the screen X120 might be up to 10%.

Table 1: Beam Non-Normalized Emittances Obtained with Simple Geometrical Method

|   | $\sigma_{X120}$, µm | $\sigma_{X109}$, µm | $\varepsilon_{X109}$, nm | $\sigma_{X111}$, µm | $\varepsilon_{X111}$, nm |
|---|---|---|---|---|---|
| **X** | 136 | 168 | 2.05 | 163 | 2.09 |
| **Y** | 87 | 300 | 3.83 | 268 | 3.79 |

The first model-dependent orbit correction relative to the quadrupoles' magnetic axes was done after the initial setup of FAST was completed for the 2016 run. Initial and final misalignments are presented in Figure 3 and Figure 4 with corrected trajectory offsets of less than 500 µm. Later, several smaller corrections were performed that showed good trajectory stability.

In theory, presented technique of trajectory correction should work for any focusing element, including two superconductive capture cavities located after the gun. According to the linear model presented in [3], RF voltage and phase can both be varied to change focusing inside a cavity. In practice, it is much easier to alter the RF phase. Unfortunately, attempts to measure the trajectory in resonators, which were done before the linear lattice calibration, were unsatisfactory. There was no numerical agreement between the applied trajectory bumps and the reconstructed responses in the capture cavities. There are several possible explanations for observed inconsistency:

- Big uncertainties in trajectory responses to the RF phase variations
- Discrepancy between actual and model RF phase and voltage
- Some models predict strong nonlinear RF field around couplers on both ends of the capture cavities that might make the used method inapplicable

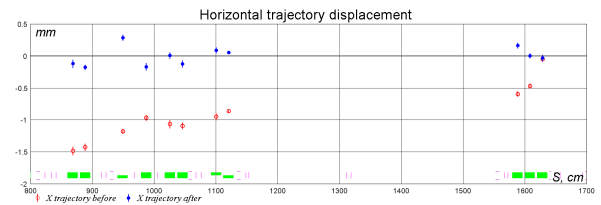

Figure 3: Correction of the horizontal trajectory offsets in quadrupoles

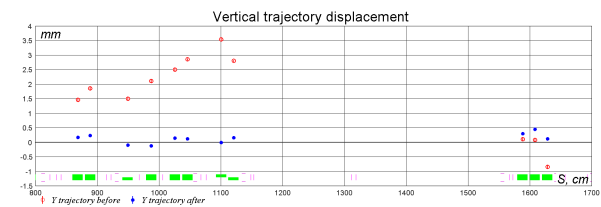

Figure 4: Correction of the horizontal trajectory offsets in quadrupoles

Lattice correction was done for the real experimental setup of capture cavities and quadrupoles. First, parameters of focusing elements were fit by analyzing trajectory responses to the dipole correctors. Then, initial conditions were reconstructed with the help of beam second moments measured at several beam profile monitors located along the FAST.

Figure 5 and Figure 6 show the difference of the trajectory responses calculated with the initial and the fitted models along with measured points on an example of the vertical and horizontal dipole correctors H101 and V101. Figure 7 and Figure 8 illustrate the difference in beam envelopes calculated using the model with quadrupole's gradients derived from set currents and from the fitted model for the same initial conditions. Differences between gradients of quadrupoles derived from set currents and taken from the fitted model are shown in Table 2. On average, all quadrupoles show the same tendency to have bigger fields than anticipated by 2.6 % with standard deviation of 1.2 %.

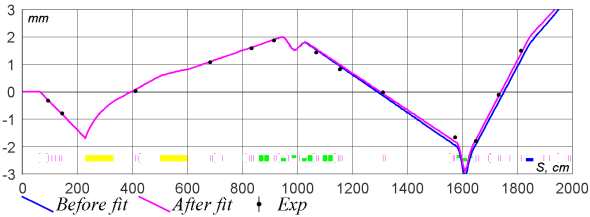

Figure 5: Trajectory response to the horizontal corrector H101 before and after fit of model parameters compared to experimental measurements

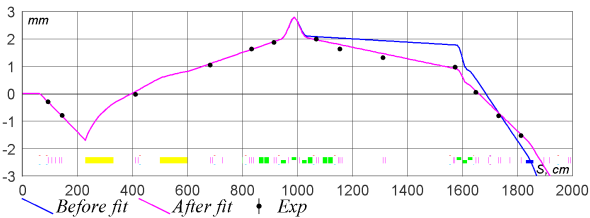

Figure 6: Trajectory response to the vertical corrector V101 before and after fit of model parameters compared to experimental measurements

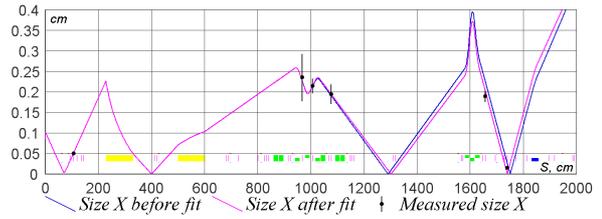

Figure 7: Horizontal beam envelope along FAST before and after model fit compared to the experimental measurements

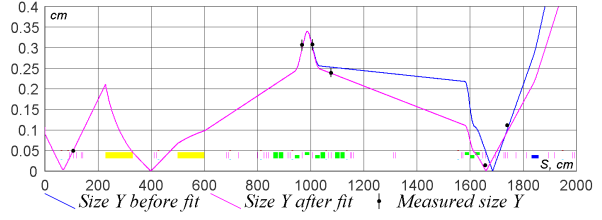

Figure 8: Vertical beam envelope along FAST before and after model fit compared to the experimental measurements

Table 2: Model and Reconstructed Gradients in Quadrupoles

| Quad | $G_{set}$, kG/cm | $G_{fit}$, kG/cm | Err, % |
|---|---|---|---|
| Q108 | 0.0809 | 0.0816 | 0.93 |
| Q109 | -0.1373 | -0.1422 | 3.57 |
| Q110 | 0.0808 | 0.0821 | 1.57 |
| Q118 | -0.2307 | -0.2371 | 2.76 |
| Q119 | 0.3405 | 0.3514 | 3.21 |
| Q120 | -0.1184 | -0.1229 | 3.82 |

## CONCLUSION

Developed algorithms allowed to correct the trajectory to within 0.5 mm relative to the magnetic axes of quadrupoles. More studies are needed to properly apply the same technique to the alignment of the trajectory in the capture cavities. A realistic model of the FAST linear accelerator was created based on the trajectory responses to the dipole correctors and the beam sizes along the accelerator. Further automation of both trajectory and lattice correction scripts is needed for routine use during FAST setup for a specific experiment.

## REFERENCES

[1] D. Edstrom, Jr. *et al.*, *The 50-MeV run in the FAST electron accelerator*, NAPAC16, Chicago, IL, October 2016, paper TUPOA19, this conference.

[2] A. Burov, S. Nagaitsev, A. Shemyakin, and Ya. Derbenev, *Optical principles of beam transport for relativistic electron cooling*, Phys. Rev. ST Accel. Beams 3, 094002, 2000.

[3] E. E. Chambers, *Particle motion in a standing wave linear accelerator*, HEPL 570, Oct. 1968.